# THE IMPACT OF EMPLOYMENT WEB SITES' TRAFFIC ON UNEMPLOYMENT: A CROSS COUNTRY COMPARISON

**M. Lütfi ARSLAN**
Business Management Dept.
Istanbul Medeniyet University
Asst. Prof. Dr.

lutfi.arslan@medeniyet.edu.tr

**Sadi Evren SEKER**
Business Management Dept.
Istanbul Medeniyet University
Asst. Prof. Dr.
academic@sadievrenseker.com

─Abstract─

Although employment web sites have recently become the main source for recruitment and selection process, the relation between those sites and unemployment rates is seldom addressed. Deriving data from 32 countries and 427 web sites, this study explores the correlation between unemployment rates of European countries and the attractiveness of country specific employment web sites. It also compares the changes in unemployment rates and traffic on all the aforementioned web sites. The results showed that there is a strong correlation between web sites traffic and unemployment rates.

**Key Words :** Unemployment, Job Market, Cross-Country Market, Data Mining, Time Series Analysis.

**JEL Classification: M15, M16, M5**

**Acknowledgement**

This study is supported by Istanbul Medeniyet University, research projects department.

## 1. INTRODUCTION

This study is built on web statistics of employment web sites and the unemployment rates of the European countries. The unemployment rates are collected from the public data published by Eurostat and a job search site, European Youth Portal, which is founded and operated by European Commision. The job search sites are mostly the web sites holding public information about the job market structure of the origin country as well as the practical informations like preparing CV or finding internship etc. Unfortunately, some of the web sites on the portal are prepared in English, which makes them relatively less attractive for the local jobs while they are more visible and attracted to the international job seekers.

**Figure-1. Overview of Study**

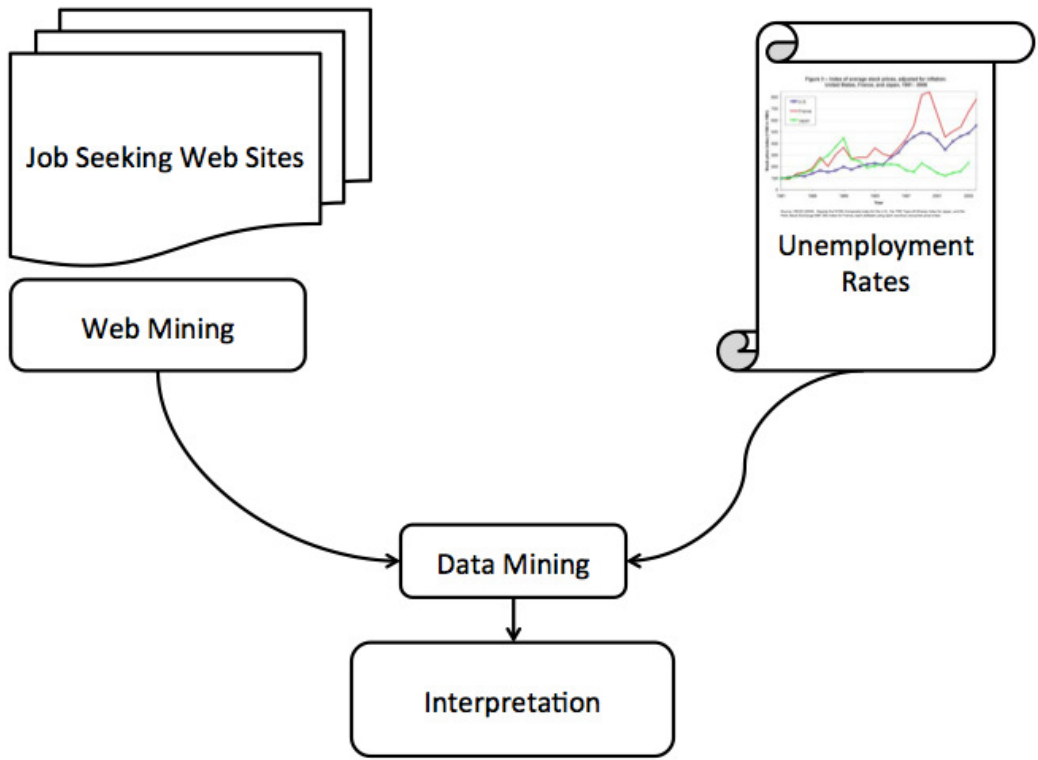

Figure 1 demonstrates the overview of the data collection and correlation steps. While, the job seeking web sites are studied via the web mining methods and statistical data is gathered from the web tools, the unemployment rates are considered as the second input to the correlation algorithms. Then, the results gathered from the data mining step are interpreted. Please note that the above diagram holds the study for only one of the countries. The study is repeated for 32 EU Countries (Members and Candidates) and also the overall data is processed for the whole European union.

This paper starts with the problem statement and the background of the unemployment rate and job market studies. We also propose the data mining methodology and the evaluation of data sets on the following sections. Finally the correlation between the job searching web sites and unemployment rates are interpreted and also the correlation between some countries are underlined.

## 2. BACKGROUND

Since employment web sites have recently become focal points of job seekers, their quantity and impact have increased. These websites are not only first venue for job seekers to attract (Carlson,Connerley, and Mecham,2002), but also economic and convenient tool of recruiting for companies, reducing the cost per hire (Galanaki,2002). Delving into the question of how the Internet has impacted job search behavior, Stevenson (2008) found that job search behavior became more extensive and search methods used by the unemployed varied. Likewise Bagues and Labini (2007) found a similar correlation between Internet data and labor market outcomes.

The relation between data flow of the Internet and structural changes in the labor market have been subject to numerous studies. Especially search engine data has been used as an indicator of economy. For instance Askitas and Zimmermann (2009), using Google search data, found correlations between keywords used in searches and unemployment rates of Germany. While D'Amuri (2009) used Google Index to predict quarterly unemployment rates of Italy, McLaren and Shanbhogue (2011) analyzed the relation between search behaviors of users and labor markets in UK. A study by Ettredge, Gerdes and Karuga found positive and significant association between job search data and the official unemployment data of US. Finally, Choi and Varian (2011) showed that search data could be used to forecast automobile sales, unemployment claims, travel destination planning, and consumer confidence.

To our best knowledge, there is no study to look for a relation between website traffic and unemployment. This study is first to do so. Although the relation between any data flow of the Internet and labor market outcomes should be treated cautiously, this study as a cross-country comparison gives a meaningful result that there could be a correlation between web site traffic and unemployment rates.

## 3. METHODOLOGY

This chapter covers the details of web mining and data mining parts of the project. In the web mining phase, the web sites are crawled manually and the required information is collected. Also in the data mining phase, the gaussian processes methodology has been deployed in order to find out the correlation between the unemployment rates and the job related governmental web sites.

### 3.1. Web Mining

During the web mining phase (Seker, et al 2013), the web sites holding the employment information are crawled manually and the information related to the web sites are collected from independent third party organizations like alexa.com, google trends or web site ranking sites. The job finding sites are collected from European Youth Portal, prepared and published by European Union[1]. The unemployment rates are collected from the Google Public Data and the countries are limited with the European countries only.

After collecting the data from several web resources, the collected information is normalized in order to integrate. Finally all the collected data is summarized in a single value for each of the web sites, wehere the second column of the data set is holding the unemployment rate of the country of the web site.

Because of dirty data after collection, the imputation technique, list wise deletion has been applied over the data set. In this imputation technic, any dirty data entry is deleted completely. For example if we can reach the web site, its ranking in alexa and its google trend but its web site traffic the whole information is considered as dirty and we do not take this record into consideration.

### 3.2. Data Mining

Studies about supervised learning can be categorized into two groups. The first group of studies is the regression method where the data is continuous and optimization of study provides a closer prediction most of the time. The second group of studies is the classification where the prediction ends up with a discrete set of classes. (Ocak and Seker,2013)

---

[1] Full path of the portal is :
http://europa.eu/youth/working/finding_a_job/index_he_en.html

Gaussian Processes (GP) is a prediction method applied on continuous data types and it can be considered as a Gaussian distribution over the function domain. The function domain can be treated by two perspective, parametric or non-parametric statistics. This study is built mainly on parametric statistics where it is attempts optimize the known number of function parameters. Learning from training data sets modifies the distribution and adapts the ensemble of parameters.

In a more general definition, Gaussian processes for regression (GPR) can be defined as a supervised learning method where the Gaussian probability distribution (GPD) is applied over the random variables of the function.

$$P(y) = \frac{1}{\sigma\sqrt{2\pi}} e^{-\frac{(y-\mu)^2}{2\sigma^2}} \qquad (2)$$

Equation 2, demonstrates the classical GPD function (Fig. 9), where y is a continuous variable, σ is the standard deviation and μ is the mean of the distribution (or can be considered as the mode or median).

**Figure-2. The generic view of distribution.**

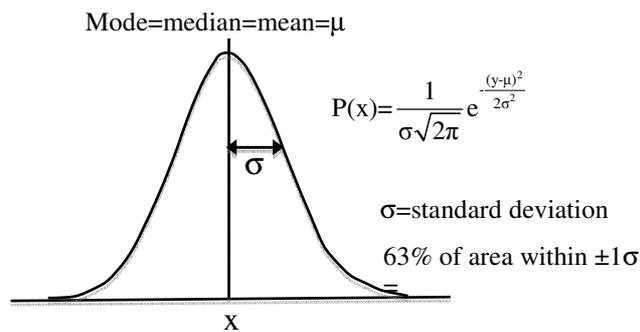

Any GPR application requires a function of random variable x and the covariance matrix C(x, x□). A function y(x) indicates the Gaussian distribution value of given x variable. The covariance can be imagined as a function that expresses the expected covariance between the values of function y at the points x and x□ (Stein, 1999)

$$C(x, x□) = E[(u(x) - m(x))(u(x□) - m(x□))] \qquad (3)$$

Equation 3 is given to show the calculation of covariance matrix $C(x, x')$, where $u(x)$ is a random process and $m(x)$ is the mean function of variable x. From the definition, it can be concluded that $m(x) = E[u(x)]$.

The purpose of regression by Gaussian Processes is the prediction of $t_{N+1}$ by processing the vector $t_N$. In this case, if the covariance matrix of $t_{N+1}$ is symbolized by $C_{N+1}$, the matrix will be in the form of Equation 4. (Ocak and Seker, 2013)

$$C_{N+1} \equiv \begin{bmatrix} [C_N] & [k] \\ [k^T] & [k] \end{bmatrix} \qquad (4)$$

Please note that, the dimensions of $C_{N+1}$ are $(N+1) \times (N+1)$

The function $y(x)$ can be expressed as in equation (5).

$$y(x) = f(x) + Z(x) \qquad (5)$$

Where Z is a stochastic process over x and f is a multivariate linear regression model as displayed in equation (6):

$$f(x) = F(x) \cdot B = \sum_{i=0}^{n} b_i \cdot f_i(x) \qquad (6)$$

Where B is the regression parameter and B = {b0, b1, … bn} and F(x) = |fo(x), f1(x), … fn(x)|.

During this research, the Gaussian Processes are utilized as a correlation model with Bayesian Formalism. The covariance effect over x can be demonstrated as in equation (7) (Ocak and Seker, 2013).

$$\sum\nolimits_{sj \cdot k} = \text{Cov}[Z(x^{(j)}), Z(x^{(k)})]_{j,k=1,\dots,N_S} = \sigma^2 \times \rho(x^{(j)}, x^{(k)})_{j,k=1,\dots,N_S} \qquad (7)$$

In equation (7), the nxn covariance matrix is utilized and the result is equal to interpolation between the xi and xk.

The Gaussian Processes formalizes the correlation between xi and xk as in equation (8):

$$\rho(x^{(j)}, x^{(k)}) = \prod_{i=1}^{n} \rho(x_i^{(j)} - x_i^{(k)}) = \exp\left\{-\sum_{i=0}^{n} \frac{\left\|x_i^{(j)} - x_i^{(k)}\right\|^2}{\theta_i}\right\} \quad (8)$$

One crucial parameter for the above formulation is the theta value, which indicates the hyper-parameter and can be interpreted as the correlation length.

## 4. DATA SET

The data set collected after the web mining phase, is normalized and collected into a single output. At the end of processing web mined data we end up with two seperate columns. In the first column we collect the normalized score for the web traffic and on the second column we collect the unemployment rate of the country where the site is orginated.

Properties of the data set is provided in Table 1.

| **Property** | **Value** |
|---|---|
| Number of Web Sites | 427 |
| Number of Web Sites after Imputation | 382 |
| Average Unemployment rate among the countries of the web sites | 7.7 |
| Average Unemployment rate of the European Union | 8.7 |
| Standard Deviation of unemployment rate of web page | 3.29E-14 |
| Average ranking of the web page | 2,509,766 |
| Standard Deviation of web page ranking | 5,498,178 |
| **Correlation Rate** | **54.49%** |
| RMSE | 0.504 |
| RAE | 0.981 |

The RMSE (Ocak and Seker,2013) stands for root mean square error and RAE (Ocak and Seker,2013) stands for root absolute error rates which are given to indicate the error rates ont he given data mining study.

## 5. CONCLUSION

The correlation rate indicates that there is a close correlation between the web site trend and the unemployment rate of the country where the web page is originated. It is also a two way operator where the correlation can be interpreted as a predictability. So from the unemployment rate of the country the web traffic of the job oriented web sites can be predicted or vice versa.

Deriving data from 32 countries and 427 web sites, this study explores the correlation between unemployment rates of European countries and the attractiveness of country specific employment web sites. It also compares the changes in unemployment rates and traffic on all the aforementioned web sites. The results showed that there is a strong correlation between web sites traffic and unemployment rates.